\documentstyle[preprint,aps]{revtex}

\begin{document}

\draft

\preprint{LA-UR-96-0782}

\title{
Realistic Expanding Source Model for Invariant One-Particle
Multiplicity Distributions and Two-Particle Correlations in
Relativistic Heavy-Ion Collisions}

\author
  {Scott Chapman and J. Rayford Nix}

\address{
  Theoretical Division, Los Alamos National Laboratory,
  Los Alamos, NM  87545, USA
}

\date{May 1, 1996}

\maketitle

\begin{abstract}
\\[-25pt]
We present a realistic expanding source model with nine parameters
that are necessary and sufficient to describe the main physics
occuring during hydrodynamical freezeout of the excited hadronic
matter produced in relativistic heavy-ion collisions.  As a first test
of the model, we compare it to data from central Si + Au collisions at
$p_{\rm lab}/A=14.6$ GeV/$c$ measured in experiment E-802 at the
AGS\@.  An overall $\chi^2$ per degree of freedom of 1.055 is achieved
for a fit to 1416 data points involving invariant $\pi^+$, $\pi^-$,
$K^+$, and $K^-$ one-particle multiplicity distributions and $\pi^+$
and $K^+$ two-particle correlations.  The 99\%-confidence region
of parameter space is identified, leading to one-dimensional error
estimates on the nine fitted parameters and other calculated physical
quantities.  Three of the most important results are the freezeout
temperature, longitudinal proper time, and baryon density along the
symmetry axis.  For these we find values of 92.9 $\pm$ 4.4 MeV, 
8.2 $\pm$ 2.2~fm/$c$, and 0.0222~$^{+0.0096}_{-0.0069}$~fm$^{-3}$,
respectively.
\end{abstract} 


\narrowtext

\section{Introduction}
It is a widely accepted theory that if nuclear matter attains a high
enough energy density, it will undergo a phase transition from normal
hadronic matter into a quark-gluon plasma (QGP)
\cite{qgp1,qgp2,qgp3,qgp4}.  Since the discovery of such a QGP would
represent a significant advancement in the fundamental understanding
of nuclear interactions, there are a number of relativistic heavy-ion
experiments both currently running and being planned which hope to
test this theory.  Unfortunately, if a QGP is formed in the
laboratory, its quick expansion and cooling will cause it to
transition back into normal hadronic matter long before anything can
be detected.  Thus, any signals for the prior existence of a QGP will
necessarily be subtle and indirect.

In order to work backwards from the final observed state of the
detected hadrons to an earlier state which may or may not have
included a QGP, it is necessary to use a reliable transport model.
One approach which has been quite successful in the past is to treat
the expanding nuclear matter as a hydrodynamical fluid.  This fluid is
very hot and dense immediately after the collision, but with time it
expands and cools.  When some criterion is met (e.g., falling below a
certain temperature or density), it is assumed that the fluid
``freezes out'' and becomes a collection of non-interacting,
free-streaming hadrons.  The freezeout hypersurface is thus some
three-dimensional surface which separates hydrodynamically interacting
nuclear fluid from free-streaming hadrons.  According to this picture,
when these hadrons are observed in detectors, their distributions and
correlations contain information about the temperature, expansion
velocities, chemical potentials, size, and shape of the fluid during
freezeout.

The purpose of this paper is to present a physically reasonable
parametrization of the freezeout process, and then to find the best
values for the freezeout parameters by comparing theoretical
distributions and correlations to experimental data through a
minimization of $\chi^2$.  This approach is somewhat different from a
standard nuclear hydrodynamical approach, in which some equation of
state must be assumed in order to determine how the fluid evolves from
its initial condition to its final freezeout \cite{hylander}.  One
problem with standard nuclear hydrodynamics is that the formidable
computations involved make a minimization of $\chi^2$ impractical, so
even when the agreement with experiment is quite good, one can never
be sure that the {\em best} point in the infinite-dimensional space of
all possible initial conditions, equations of state, and freezeout
criteria has been found.  Our more limited goal is to tackle just the
problem of determining the properties of the system during freezeout.

We begin by reviewing the Wigner-function formulation of
hydrodynamical freezeout and defining nine parameters that are
necessary and sufficient to properly describe the gross properties of
the source during freezeout.  Although in Sec.~II.B we use the
language of hydrodynamical evolution (e.g., rarefaction waves and
cooling) to motivate our approach, it should be noted that our
calculations are actually concerned only with freezeout --- not with
the hydrodynamical evolution which might have led to it.  Section II.D
then includes a short explanation of how resonance decays are taken
into consideration.  Once the model is defined, Sec.~III outlines our
general method for constructing $\chi^2$, determining the goodness of
the fit, and estimating uncertainties in the model parameters.  With
these tools in hand, we compare our nine-parameter model to data from
central Si + Au collisions at $p_{\rm lab}/A=14.6$ GeV/$c$, measured
in experiment E-802 at the Brookhaven Alternating Gradient Synchrotron
(AGS)
\cite{e802,dbase,vince}.  The 1416 data points used consist of
invariant $\pi^+$, $\pi^-$, $K^+$, and $K^-$ one-particle multiplicity
distribution measurements as well as $\pi^+$ and $K^+$ two-particle
correlation measurements.  To our knowledge, this paper represents the
first attempt to simultaneously find the best fit to one-particle
distribution and three-dimensional two-particle correlation data with
a single expanding source model.  We found that the fits converged
rapidly and consistently, yielding an overall $\chi^2$ per degree of
freedom of 1.055.

%%%%%%%%%%%%%%%%%%%%%%%%%%%%%%%%%%%%%%%%%%%%%%%%%%%%%%%%%%%%%%%%%
\section{Details of the Model}\label{s1}
%%%%%%%%%%%%%%%%%%%%%%%%%%%%%%%%%%%%%%%%%%%%%%%%%%%%%%%%%%%%%%%%%

\subsection{Wigner Function Formulation}

The Wigner function for particles of type $\alpha$ with spin
$J_\alpha$ coming directly from a hydrodynamical system involving a
sharp three-di\-men\-sio\-nal freezeout hypersurface is \cite{marb}
\begin{equation}
   S^{\rm dir}_\alpha(x,p) = \frac{2J_\alpha+1}{(2\pi)^3}
   \frac{p{\cdot}n(x)}{\exp\{[p{\cdot}u(x)-\mu_\alpha(x)]/T(x)\} 
   \mp1 } \;,
 \label{1.1}
\end{equation}
where the $-$ ($+$) sign is for bosons (fermions).  The quantities
$u^\mu(x)$, $T(x)$, $\mu_\alpha(x)$, and
\begin{equation}
   n_\mu(x)=
   \int_\Sigma d^3\sigma_\mu(x^\prime)\,\delta^{(4)}(x-x^\prime)
 \label{1.2}
\end{equation}
denote the local hydrodynamical flow velocity, temperature, chemical
potential, and normal-pointing freezeout hypersurface element,
respectively.  Throughout the paper we use units in which
$\hbar=c=k=1$, where $\hbar$ is Planck's constant divided by $2\pi$,
$c$ is the speed of light, and $k$ is the Boltzmann constant (except
in the figures and tables, where we reinsert $c$).  Integrating the
direct Wigner function over spacetime generates the Cooper-Frye
formula for the one-particle distribution \cite{coop}:
\begin{equation}
P_\alpha^{\rm dir}({\bf p}) = \int d^4x\, S^{\rm dir}_\alpha(x,p) 
= \frac{2J_\alpha+1}{(2\pi)^3}\int_\Sigma d^3\sigma_\mu
   \frac{p^\mu}{\exp[(p{\cdot}u(x)-\mu_\alpha(x))/T(x)] 
   \mp1 }\;.
\label{1.3}
\end{equation}
The subscript $\Sigma$ on the integral denotes the limits to the
hypersurface for a finite-sized system.  Because the observed
particles are on mass shell, $P_\alpha^{\rm dir}$ depends only on the
three-vector ${\bf p}$ rather than on the four-vector $p$.

In addition to particles coming directly from the freezeout surface,
there are also some which come from the decay of resonances.  The
total Wigner function for particle $\alpha$ is then comprised of two
parts:
\begin{equation}
S_\alpha(x,p) = S_\alpha^{\rm dir}(x,p) + S_{{\rm res}\to \alpha}(x,p)\;,
\label{1.4}
\end{equation}
where the second term is determined by the direct Wigner functions of
the contributing resonances (see Sec.~II.D).  The total observed
multiplicity distribution for particle $\alpha$ is
\begin{equation}
P_\alpha({\bf p}) = \int d^4x\, S_\alpha(x,p)\;.
\label{1.5}
\end{equation}

The correlation function for two particles of type $\alpha$ with
momenta ${\bf p}_1$ and ${\bf p}_2$ can similarly be expressed in
terms of the Wigner function \cite{shur,pratt,plum,chap1,bertsch1}
\begin{equation}
C_\alpha({\bf q},{\bf K}) = 1 \pm \lambda_\alpha
\frac{\left|\int d^4x S_\alpha(x,K)\exp(iq{\cdot}x)\right|^2}
{P_\alpha({\bf p}_1)\,P_\alpha({\bf p}_2)}\;,
\label{1.6}
\end{equation}
where the deviation from unity of the parameter $\lambda_\alpha$
measures the amount of coherent production of particles of type
$\alpha$.  Although the on-shell momenta of the two particles is
completely specified by the six momentum components in ${\bf
K}={\textstyle\frac{1}{2}}({\bf p}_1+{\bf p}_2)$ and ${\bf q}={\bf
p}_1-{\bf p}_2$, it is nevertheless notationally convenient to make
the full off-shell four-vector definitions
$K={\textstyle\frac{1}{2}}(p_1+p_2)$ and $q=p_1-p_2$.  Since both the
one-particle distribution and the two-particle correlation function
are determined by the Wigner function, we need only find a suitable
parametrization of this function in order to compare a hydrodynamical
model to these data.

\subsection{Definition of the Model Parameters}
Our model is applicable to nearly central collisions of
ultrarelativistic nuclei.  For large sets of many nearly central
collisions, the data should be azimuthally symmetric, so we assume
azimuthal symmetry in our model.  Immediately after the collision, we
assume the formation of a hot, dense source which moves with some
velocity $v_{\rm s} =\tanh y_{\rm s}$ relative to the lab while it
expands and cools in its own rest frame.  If the incoming nuclei are
relativistic enough in the source frame, their strong Lorentz
contraction makes their thickness in the beam ($z$) direction
negligible, so it should be a good approximation to assume that the
collision took place on a single plane at $t=z=0$.  Assuming also that
the longitudinal flow velocities subsequently imparted to each bit of
the nuclear fluid remain constant throughout the expansion, these
velocities take the simple form first suggested in
\cite{coop2,bjork}, namely
\begin{equation}
\beta_z(\eta) = \frac{z}{t} = \tanh\eta\;,
\label{2.4}
\end{equation}
where $\eta=\tanh^{-1}(z/t)$ is the spacetime rapidity of that bit of
fluid in the source frame.  We will show shortly that this
flow profile leads to a longitudinally boost-invariant local energy.

Unlike in the longitudinal direction, there is no initial motion in
the transverse direction.  After the collision, however, rarefaction
waves work their way radially inward, causing the matter to accelerate
transversally outward.  The resulting three-dimensional expansion
causes the fluid to cool until eventually a low enough temperature is
reached so that the matter effectively stops interacting and ``freezes
out.''  We consider a model in which both the temperature $T$ and
chemical potential are constant at freezeout.  For the latter, we
define
\begin{equation}
\mu_\alpha = B_\alpha\mu_{\rm b} 
+ S_\alpha\mu_{\rm s} + I_\alpha\mu_{\rm i}\;.
\label{2.1}
\end{equation}
Here $B_\alpha$, $S_\alpha$, and $I_\alpha$ are the baryon,
strangeness, and isospin numbers of particle type $\alpha$, while
$\mu_{\rm b}$, $\mu_{\rm s}$, and $\mu_{\rm i}$ are the corresponding
chemical potentials.

Although there are many possible ways to parametrize the radial flow
at freezeout, the actual profile chosen may not be nearly as important
as the average transverse velocity of the profile \cite{shin}.  Recent
hydrodynamical studies have obtained transverse flow profiles which
are relatively linear in $\rho=\sqrt{x^2+y^2}$ out to a certain
radius, outside of which they drop off quickly
\cite{schlei,heinz,schn}.  For simplicity, we assume a linear profile,
and to preserve boost-invariance, we follow \cite{schn,chap2,csor} by
defining it to be independent of $z$ and $t$ in the longitudinally
comoving frame of the source.  In other words, we parametrize the
total flow velocity of the system in the source frame by
\begin{equation}
u^\mu(x) = \gamma_\rho\biggl(\;\cosh\eta,\;
\beta_\rho\cos\phi,\;\beta_\rho\sin\phi,\;
\sinh\eta\;\biggr)\;,
\label{2.2}
\end{equation}
where $\gamma_\rho=1/\sqrt{1-{\beta_\rho}^2}$, with
\begin{equation}
\beta_\rho(\rho) = v_{\rm t}\left(\frac{\rho}{R}\right)\;.
\label{2.3}
\end{equation}
Here $R$ is the maximum transverse radius of the source and $v_{\rm
t}$ is the magnitude of the transverse velocity of the fluid at
$\rho=R$.  Note that $\beta_\rho$ is the flow velocity in the
longitudinally comoving frame, but that the transverse component of
the total flow velocity in the source frame is $\beta_\rho/\cosh\eta$.

That the flow profile is in fact boost-invariant can be most easily
seen by first rewriting the source-frame particle four-momentum in the
form
\begin{equation}
p^\mu = \left(\;m_{\rm t}\cosh (y_{\rm p}-y_s),\;{\bf p}_\perp,
\;m_{\rm t}\sinh (y_{\rm p}-y_s)\;\right)\;,
\label{2.5}
\end{equation}
where $m_{\rm t}=\sqrt{E^2-{p_z}^2}$ is the ``transverse mass,''
$y_{\rm p}=\tanh^{-1}(p_z/E)$ is the rapidity of the particle in the
lab frame, and ${\bf p}_\perp$ is the transverse momentum two-vector.
Throughout this paper, we will use the subscript $\perp$ to denote the
vector made from the two transverse components of a four-vector.  Note
that the quantity
\begin{equation}
p{\cdot}u(x) = \gamma_\rho m_{\rm t}
\cosh\left[(y_{\rm p}-y_{\rm s})-\eta\right] 
- {\bf p}_\perp{\cdot}{\bf u}_\perp(\rho)
\label{2.6}
\end{equation}
depends on the rapidity of the particle and the spacetime rapidity of
the source only through their difference.  Since boosting to a frame
moving with longitudinal velocity $U$ relative to the source frame can
be done by subtracting $\tanh^{-1}(U)$ from both $(y_{\rm p}-y_{\rm
s})$ and $\eta$, the difference of these quantities is boost
invariant.

In keeping with the boost-invariant profile, we assume that freezeout
along the $\rho=0$ symmetry axis of the source occurs at a constant
proper time \cite{bjork}.  Due to transverse expansion effects,
however, freezeout may occur sooner for matter with $\rho\ne 0$.
These assumptions are incorporated into the following equation
describing the freezeout hypersurface:
\begin{equation}
\frac{t^2 -  z^2}{1+\alpha_{\rm t}(\rho/R)^2}
= {\tau_{\rm f}}^2 ={\rm const.}\;,
\label{2.7}
\end{equation}
where $\alpha_{\rm t}$ parametrizes the radial behavior of the
freezeout process.  At a given constant slice in $z$, for
$-1<\alpha_{\rm t}<0$, freezeout proceeds radially from outside to
inside.  For example, freezeout for the $z=0$ slice begins on the
outside ($\rho=R$) at time $t=t_1=\tau_{\rm f}\sqrt{1+\alpha_{\rm t}}$
and continues until the inside ($\rho=0$) freezes out at time
$t=t_2=\tau_{\rm f}$.  The case $\alpha_{\rm t}>0$ corresponds to the
less-likely possibility of freezeout proceeding radially from the
inside to the outside, while $\alpha_{\rm t}=0$ represents a freezeout
which occurs at the same time for all points with a given $z$.  The
temporal duration of freezeout for the $z=0$ slice is just given by
\begin{equation}
\Delta t(z=0) = \left|t_2 - t_1\right|
= \tau_{\rm f}\left|1 - \sqrt{1+\alpha_{\rm t}}\right|\;.
\label{2.7a}
\end{equation}

To derive the prefactor $p{\cdot}n(x)$ for the spacelike hypersurface
defined by Eq.~(\ref{2.7}), it is most convenient to use the spacelike
variables $\eta$, $x$, and $y$.  We have \cite{schn}
\begin{equation}
d^3\sigma_\mu(x) = \epsilon_{\mu\nu\alpha\beta}\frac{dX^\nu}{d\eta}
\frac{dX^\alpha}{dx}\frac{dX^\beta}{dy}\,d\eta dxdy\;,
\label{2.8}
\end{equation}
where the coordinate vector on the hypersurface is given by
\begin{equation}
X^\mu = \left(\;\tau_{\rm f}\sqrt{1+\alpha_{\rm t}(\rho/R)^2}
\cosh\eta,\;x,\;y,
\;\tau_{\rm f}\sqrt{1+\alpha_{\rm t}(\rho/R)^2}\sinh\eta\;\right)\;.
\label{2.9}
\end{equation}
Doing the algebra, we find
\begin{equation}
p{\cdot}d^3\sigma(x) = \tau_{\rm f}\left\{
\sqrt{1+\alpha_{\rm t}(\rho/R)^2}\,m_{\rm t}\cosh\left[(y_{\rm
p}-y_{\rm s})-\eta\right] 
-\alpha_{\rm t} {\bf p}_\perp{\cdot}{\bf x}_\perp
\tau_{\rm f}/R^2\right\}d\eta dxdy\;,
\label{2.10}
\end{equation}
where ${\bf x}_\perp=(x,y)$.  Again, since the only $(y_{\rm p}-y_{\rm
s})$ and $\eta$ dependencies come through the difference $[(y_{\rm
p}-y_{\rm s})-\eta]$, the above expression is longitudinally boost
invariant.

Since we are interested in realistic finite systems, it is necessary
to put some spacelike limits on the hypersurface.  We have already
mentioned the maximum transverse radius $R$; we also assume the
existence of a maximum longitudinal radius $z_3=\tau_{\rm
f}\sinh\eta_0$ which is achieved by the source at time $t_3=\tau_{\rm
f}\cosh\eta_0$.  Because colliding nuclei have more matter in the
center ($\rho=0$) than on the outside, we take our source to be
spheroidal in $\rho$ and $\eta$ (as opposed to cylindrical, for
example).  In other words, on the hypersurface $\Sigma$, the spacelike
coordinates $\rho$ and $\eta$ satisfy the inequality
\begin{equation}
\frac{\rho^2}{R^2} + \frac{\eta^2}{{\eta_0}^2} \le 1\;.
\label{2.11}
\end{equation}
Since in the above equation, $\eta$ appears alone and not in
combination with $(y_{\rm p}-y_{\rm s})$, these limits break boost
invariance.  Using Eq.~(\ref{1.2}) and the definition
$\tau=\sqrt{t^2-z^2}$, we obtain in the source frame
\begin{eqnarray}
p{\cdot}n(x) &=& 
\left\{m_{\rm t}\cosh\left[(y_{\rm p}-y_{\rm s})-\eta\right]
-\frac{\alpha_{\rm t} {\bf p}_\perp{\cdot}{\bf x}_\perp\tau_{\rm f}}
{R^2\sqrt{1+\alpha_{\rm t}(\rho/R)^2}}\right\}
\nonumber \\	
&\times&\delta
\left(\tau-\tau_{\rm f}\sqrt{1+\alpha_{\rm t}(\rho/R)^2}\right)
\theta\left(1-(\rho/R)^2-(\eta/\eta_0)^2\right)\;.
\label{2.12}
\end{eqnarray}

The freezeout hypersurface as a function of $t$, $\rho$, and $z$ shown
in Fig.~1 corresponds to $R=8.0$ fm, $\tau_{\rm f}=8.2$ fm/$c$,
$\eta_0=1.47$, and $\alpha_{\rm t}=-0.86$, which are parameters that
we determine in Sec.~IV.B for the reaction considered here
\cite{e802,dbase,vince}.  As mentioned before, since $\alpha_{\rm t}$
is negative, freezeout begins on the outside at $\rho=R$ and works its
way in, reaching the center at $t=t_2=\tau_{\rm f}$.  Figure 2 shows
an illustration of the hydrodynamical fluid for seven different
instances in source-frame time.  The inner surfaces at each time are
actually freezing out, while the outer end caps are the boundaries of
fluid which will freeze out later.  Since the end caps are not yet on
the freezeout hypersurface, their exact shapes are not actually
determined by our model and are shown only to illustrate the finite
nature of the source.  By time $t=\tau_{\rm f}$, freezeout has worked
its way to the center, so for later times the source becomes two
separated receding fireballs of continually decreasing size.

For symmetric projectile-target collisions, the rapidity $y_{\rm s}$
of the source frame relative to the lab is given simply by the average
of the projectile and target rapidities.  For asymmetric collisions,
however, the precise value of $y_{\rm s}$ depends on how many
``participant'' nucleons in each nucleus collide to form the
hydrodynamical source.  The center of mass of the source is just the
center of mass of the incoming participants rather than the total
center of mass of all of the incoming nucleons (participants $+$
spectators).  Given the masses of the incoming nuclei, $y_{\rm s}$
could be estimated either purely on geometrical grounds or it could be
treated as another variable parameter to be fit to data.  For the
asymmetric Si + Au collisions studied in this paper, we choose the
latter approach.

Although at first we considered separate incoherence parameters for
pions and for kaons, we subsequently found that very good fits could
be obtained by setting $\lambda_K=1$ and allowing only $\lambda_\pi$
to vary as a parameter.  Moreover, two of the chemical potentials,
$\mu_{\rm s}$ and $\mu_{\rm i}$, can be determined from the remaining
nine parameters by imposing the constraints that the total strangeness
of the sum of all particles in the source vanishes and that the total
isospin per baryon is the same as that of the participants before the
collision (see next subsection).  The nine adjustable parameters of
our model can be grouped in the following way: $T$, $\mu_{\rm b}/T$,
and $\lambda_\pi$ describe intrinsic properties of the fluid; $R$,
$v_{\rm t}$, and $\alpha_{\rm t}$ describe transverse aspects of the
freezeout; and $y_{\rm s}$, $\eta_0$, and $\tau_{\rm f}$
describe longitudinal and boost-invariant aspects of the freezeout.

\subsection{Constraint Equations}

The total number of particles of type $\alpha$ which freeze out on the
hypersurface is given by
\begin{equation}
N^{\rm dir}_\alpha = \int \frac{d^3p}{E}
P^{\rm dir}_\alpha({\bf p})\;,
\label{3.1}
\end{equation}
where $P^{\rm dir}_\alpha$ is defined by Eq.~(\ref{1.3}).  Given
the numbers of each particle species, the total strangeness and
isospin per baryon of the system are simply
\begin{eqnarray}
S_{\rm tot}(\mu_{\rm s},\mu_{\rm i}) &=& \sum_\alpha S_\alpha
N^{\rm dir}_\alpha(\mu_{\rm s},\mu_{\rm i})
\nonumber \\
\frac{I_{\rm tot}(\mu_{\rm s},\mu_{\rm i})}
{B_{\rm tot}(\mu_{\rm s},\mu_{\rm i})} &=&
\frac{\sum_\alpha I_\alpha
N^{\rm dir}_\alpha(\mu_{\rm s},\mu_{\rm i})}
{\sum_\alpha B_\alpha
N^{\rm dir}_\alpha(\mu_{\rm s},\mu_{\rm i})}\;,
\label{3.2}
\end{eqnarray}
where we have suppressed all parameter dependencies except those of
$\mu_{\rm s}$ and $\mu_{\rm i}$.  The sum in $\alpha$ is over all
mesons with masses below 900 MeV and all baryons with masses below
1410 MeV\@.  For particles of a given mass, all of the isospin,
baryon, and strangeness states are considered separately since the
different chemical potential of each (see Eq.~(\ref{2.1})) leads to a
different value of $N^{\rm dir}_\alpha$.

The initial isospin per baryon of the system depends on the number of
participant protons and neutrons from each nucleus.  We define the
target proton number, nucleon number, and number of participants as
$Z_{\rm tar}$, $A_{\rm tar}$, and $B_{\rm tar}$, respectively.  Making
similar definitions for the projectile and noting that each proton
(neutron) has isospin 1/2 ($-$1/2), we find the total isospin of the
incoming participants:
\begin{equation}
I|_0 = 
\frac{B_{\rm proj}}{2A_{\rm proj}}\left[Z_{\rm proj} -
\left(A_{\rm proj}-Z_{\rm proj}\right)\right]
+\frac{B_{\rm tar}}{2A_{\rm tar}}\left[Z_{\rm tar} -
\left(A_{\rm tar}-Z_{\rm tar}\right)\right]\;.
\label{3.4}
\end{equation}
To get the isospin per baryon of the participants, we simply divide
the above equation by $B = B_{\rm proj}+B_{\rm tar}$.  The quantities
$B_{\rm proj}/B$ and $B_{\rm tar}/B$ can be determined by equating the
incoming target and projectile momenta in the participant
center-of-mass frame.  Explicitly,
\begin{equation}
m_{\rm N}B_{\rm proj}\sinh(y_{\rm proj}-y_{\rm s}) = m_{\rm N}B_{\rm
tar}\sinh(y_{\rm s}-y_{\rm tar})\;,
\label{3.5}
\end{equation}
where $y_{\rm proj}$ and $y_{\rm tar}$ are the initial projectile and
target rapidities and $m_{\rm N}$ is the nucleon mass.  Using this
relation, it is easy to show that the initial isospin per baryon of
the system is given by
\begin{equation}
\left. \frac{I}{B}\right|_0 = -\frac{1}{2} 
+ \frac{(Z_{\rm proj}/A_{\rm proj})\sinh(y_{\rm s}-y_{\rm tar})
+(Z_{\rm tar}/A_{\rm tar})\sinh(y_{\rm proj}-y_{\rm s})}
{\sinh(y_{\rm s}-y_{\rm tar}) + \sinh(y_{\rm proj}-y_{\rm s})}\;.
\label{3.6}
\end{equation}
In general, the initial isospin per baryon depends upon the parameter
$y_{\rm s}$, but for symmetric collisions (or any collision in which
$Z_{\rm proj}/A_{\rm proj} = Z_{\rm tar}/A_{\rm tar}$), $I/B|_0$ is
independent of $y_{\rm s}$.

We are now ready to write down explicitly the constraint equations for
$\mu_{\rm s}$ and $\mu_{\rm i}$:
\begin{eqnarray}
S_{\rm tot}(\mu_{\rm s},\mu_{\rm i}) &=& 0
\nonumber \\
\frac{I_{\rm tot}(\mu_{\rm s},\mu_{\rm i})}
{B_{\rm tot}(\mu_{\rm s},\mu_{\rm i})} &=&
\left. \frac{I}{B}\right|_0\;.
\label{3.7}
\end{eqnarray}
For a given set of the nine remaining parameters, these equations
allow one to find unique values of $\mu_{\rm s}$ and $\mu_{\rm i}$ as
well as their derivatives (e.g., $d\mu_{\rm s}/dT$).

Notice that the baryon number of this model reflects only the
participant baryons and is not constrained to be the same as the total
baryon number of the two incoming nuclei.  The rationale behind this
is that in many collisions there are ``spectator'' nucleons whose
evolution is not well described by a hydrodynamically expanding
source.  Since these spectators may nonetheless end up in detectors,
this model works best when it is used to fit only data from produced
particles such as mesons and antiprotons.  Of course, once the
parameters have been determined from a fit to mesons, for example, it
is always possible to compare the proton distribution predicted by the
model with the data to get an idea of how many of the measured protons
are ``participants'' and how many are ``spectators.''

It is significant that even if only meson data are considered, it is
still possible to determine the baryon chemical potential $\mu_{\rm
b}$ from the relative abundances of $K^+$'s and $K^-$'s.  For any
system which has a positive baryon number, there are more $\Lambda$'s
and $\Sigma$'s produced than $\bar{\Lambda}$'s and $\bar{\Sigma}$'s.
This leads to a net negative strangeness among all of the baryons.
From the constraint of Eq.~(\ref{3.7}), it follows that more $K^+$'s
than $K^-$'s are produced.  In other words, the difference between
$K^+$ and $K^-$ abundances provides an indirect measurement of baryon
number and hence of $\mu_{\rm b}$.

\subsection{Resonance Decays}

For single-generation decays, the resonance contribution to the Wigner
function is given by \cite{resoh,resos}
\begin{eqnarray}
S_{{\rm res}\to \alpha}(x,p) &=& 
\sum_\beta \int\frac{d^3p_\beta}{E_\beta}
\int d^4x_\beta\int_0^\infty d\tau_\beta
\Gamma_\beta\exp(-\Gamma_\beta\tau_\beta)
\nonumber \\
\;\;\;\;\;\;&\times &
\delta^4\left(x - \left[x_\beta +
\frac{\tau_\beta}{m_\beta}p_\beta\right]\right)
\Phi_{\beta\to\alpha}(p_\beta,p)S_\beta^{\rm dir}(x_\beta,p_\beta)\;,
\label{4.2}
\end{eqnarray}
where the sum in $\beta$ is over each decay channel of each resonance
that will produce a particle $\alpha$. The quantity
$\Phi_{\beta\to\alpha}(p_\beta,p)$ is the probability density that a
resonance with momentum $p_\beta$ will decay into a particle $\alpha$
with momentum $p$.  For a two-body decay $\beta\to\alpha + X$ that is
isotropic in the rest frame of the resonance,
\begin{equation}
\Phi_{\beta\to\alpha+X}(p_\beta,p) = 
\frac{b_\beta}{4\pi p_0(m_\beta;m_\alpha,m_X)}
\delta\left(E_0(m_\beta;m_\alpha,m_X) - 
\frac{p_\beta{\cdot}p}{m_\beta}\right)\;,
\label{4.3}
\end{equation}
where $b_\beta$ is the branching ratio of the particular decay channel
and
\begin{eqnarray}
E_0(m_\beta;m_\alpha,m_X) &=& \frac{1}{2m_\beta}\left({m_\beta}^2
+{m_\alpha}^2 -{m_X}^2\right) 
\nonumber \\
p_0(m_\beta;m_\alpha,m_X) &=&
\frac{1}{2m_\beta}\sqrt{\left[{m_\beta}^2-(m_\alpha+m_X)^2\right]
\left[{m_\beta}^2-(m_\alpha-m_X)^2\right]}\;.
\label{4.4}
\end{eqnarray}

A three-body decay $\beta\to\alpha + X+Y$ can be treated as a two-body
decay into $\alpha$ and a system with the combined invariant mass $M$
of particles $X$ and $Y$\@.  Since $M$ can vary from $m_X+m_Y$ to
$m_\beta-m_\alpha$, it must be integrated over with an appropriate
probability density.  In \cite{resoh,hage}, this has been shown to be
\begin{equation}
P(M) = \frac{p_0(m_\beta;m_\alpha,M)p_0(M;m_X,m_Y)}
{\int_{m_X+m_Y}^{m_\beta-m_\alpha}
dMp_0(m_\beta;m_\alpha,M)p_0(M;m_X,m_Y)}\;.
\label{4.5}
\end{equation}
Using this probability density, we have for three-body decays
\begin{equation}
\Phi_{\beta\to\alpha+X+Y}(p_\beta,p) =
\int_{m_X+m_Y}^{m_\beta-m_\alpha}dM
P(M)\Phi_{\beta\to\alpha+M}(p_\beta,p)\;.
\label{4.6}
\end{equation}

The resonances included which contribute to the charged pion
distributions are \cite{pdg}
\begin{eqnarray}
\eta &\to & \pi^+ + \pi^- + \pi^0
\nonumber \\
\eta &\to & \pi^+ + \pi^- + \gamma
\nonumber \\
\rho &\to & \pi + \pi
\nonumber \\
\omega &\to & \pi^+ + \pi^- + \pi^0
\nonumber \\
\omega &\to & \pi^+ + \pi^-
\nonumber \\
K^* &\to & K + \pi
\nonumber \\
\Delta &\to & N + \pi
\nonumber \\
\Sigma(1385) &\to & \Lambda + \pi
\nonumber \\
\Sigma(1385) &\to & \Sigma + \pi
\nonumber \\
\Lambda(1405) &\to & \Sigma + \pi\;.
\label{4.6a}
\end{eqnarray}
The $K^*$ and $\Delta$ resonances also contribute to the kaon and
nucleon distributions, respectively.

Since the model uses separate isospin, baryon, and strangeness chemical
potentials, each species of each resonance must be treated separately.
For example, there are four different channels by which negative pions
can be produced from delta decay:
\begin{eqnarray}
\Delta^-(I=-3/2,B=1) &\rightarrow & n + \pi^-\;\;,\;\;\;\;\;b = 0.994
\nonumber \\
\Delta^0(I=-1/2,B=1) &\rightarrow & p + \pi^-\;\;,\;\;\;\;\;b = 0.994/3
\nonumber \\
\bar{\Delta}^{--}(I=-3/2,B=-1) &\rightarrow & \bar{p} 
+ \pi^-\;\;,\;\;\;\;\;
b = 0.994
\nonumber \\
\bar{\Delta}^-(I=-1/2,B=-1) &\rightarrow & \bar{n} 
+ \pi^-\;\;,\;\;\;\;\;
b = 0.994/3\;.
\label{4.1}
\end{eqnarray}

\section{Determination of the Parameters by $\chi^2$ Minimization}

\subsection{Construction of $\chi^2$}

A single point $i$ of an experimentally measured one-particle
distribution for a particle of type $\alpha$ can be characterized by
the set $({\bf p}_i,P_\alpha^i,\sigma_i^{\rm stat})$, where
$\sigma_i^{\rm stat}$ is the statistical error of the measurement.
There are also systematic errors associated with these measurements,
which are usually expressed in terms of a percentage of $P_\alpha^i$
for each particle $\alpha$.  We denote the percent systematic error by
$f_\alpha$, so that the total error for point $i$ is given by
\begin{equation}
\sigma^{\rm tot}_{\alpha,i} = \sqrt{{\sigma_i^{\rm stat}}^2 
+ (f_\alpha P_\alpha^i)^2}\;.
\label{5.1}
\end{equation}
For each particle $\alpha$, we construct its contribution
${\chi_\alpha}^2$ to the total $\chi^2$ in the following way:
\begin{equation}
{\chi_\alpha}^2({\bf \theta}) = \sum_i 
\frac{\left[P_\alpha^i-P_\alpha({\bf p}_i,{\bf \theta})\right]^2}
{{\sigma^{\rm tot}_{\alpha,i}}^2}\;,
\label{5.2}
\end{equation}
where $P_\alpha({\bf p},{\bf \theta})$ is defined by Eq.~(\ref{1.5}),
${\bf \theta}$ is used to represent all of the model parameters ($T$,
$v_{\rm t}$, etc.), and the sum is over all of the measured data
points.

A similar contribution to $\chi^2$ can be constructed by comparing
two-particle correlation data to the corresponding model calculations.
The main difference with correlation measurements is that most of the
systematic errors are removed when dividing the two-particle
distribution measurement in the numerator by the product of the
one-particle distribution measurements in the denominator.  All of the
individual one- and two-particle contributions to $\chi^2$ can be
combined into an overall $\chi^2$ of the form
\begin{equation}
\chi^2({\bf \theta}) = \sum_\alpha\sum_i
\frac{\left[P_\alpha^i-P_\alpha({\bf p}_i,{\bf \theta})\right]^2}
{{\sigma_i^{\rm stat}}^2 + (f_\alpha P_\alpha^i)^2}
+ \sum_\beta\sum_i
\frac{
\left[C_\beta^i-C_\beta({\bf q}_i,{\bf K}_i,{\bf \theta})\right]^2}
{{\sigma_i}^2}\;,
\label{5.3}
\end{equation}
where we explicitly show the dependence of $\chi^2$ on the model
parameters ${\bf\theta}$.  By varying these parameters to minimize
$\chi^2$, we can find the best fit of the model to the data.  Our
programs can minimize $\chi^2$ by using either the Simplex method or
the Levenberg-Marquardt method \cite{numrec}.

\subsection{Confidence in the Overall Fit}

By the central limit theorem, it is well known that the probability
distribution of the sum of a very large number of small random
deviations almost always converges to a normal distribution.  In the
remainder of this paper, we will assume that a sufficiently large
number of measurements have been taken so that the measurement errors
are normally distributed around their true values.  With this
assumption as well as the assumption of a perfect model, the
probability density $\Pi_\nu$ of obtaining a minimum of $\chi^2$ equal
to ${\chi_{\rm min}}^2$ is given by the chi-square function
\cite{numrec,stat}
\begin{equation}
\Pi_\nu\!\left({\chi_{\rm min}}^2\right) = \frac{1}{2\Gamma(\nu/2)}
\exp(-{\chi_{\rm min}}^2/2)
\left(\frac{{\chi_{\rm min}}^2}{2}\right)^{\nu/2-1}\;,
\label{7.0}
\end{equation}
where $\nu$ is the number of degrees of freedom of the fit.  For a
model with $M$ adjustable parameters fit to $N$ data points, $\nu$ is
simply $N-M$\@.  Since the above distribution is single-peaked with a
mean of $~\nu$ and a variance of $\sqrt{2\nu}$, it follows that for a
perfect model the most probable values of ${\chi_{\rm min}}^2$ per
degree of freedom are close to one.  A value of ${\chi_{\rm
min}}^2/\nu$ much larger than one corresponds to a small probability
density and consequently leads one to question whether the model used
is a good one or whether the error bars on the data have been
underestimated.  Conversely, a value of ${\chi_{\rm min}}^2/\nu$ much
less than one seems too good to be true and leads one to question
whether the error bars on the data have been overestimated.

A more quantitative way to determine the ``goodness'' of the fit is to
integrate $\Pi_\nu(\chi^2)$ over $\chi^2$ from the ${\chi_{\rm
min}}^2$ actually found in the fit to infinity.  The resulting
function ${\cal P}_\nu({\chi_{\rm min}}^2)$ is the probability that
random measurement errors and a perfect model would lead to a minimum
of $\chi^2$ at least as big as the one actually found, ${\chi_{\rm
min}}^2$.  A fit to a ``good'' model results in a small ${\chi_{\rm
min}}^2$ and a ${\cal P}_\nu({\chi_{\rm min}}^2)$ which is greater
than some acceptable probability (e.g., 5\%).  A fit to a ``poor''
model, on the other hand, results in a large ${\chi_{\rm min}}^2$ and
a small value of ${\cal P}_\nu({\chi_{\rm min}}^2)$.  If, for example,
one obtained ${\cal P}_\nu({\chi_{\rm min}}^2)=10^{-10}$, it would be
very difficult to believe that the large value of ${\chi_{\rm min}}^2$
giving rise to that small probability had come about purely by way of
random measurement errors; it is far more likely that there is
something seriously wrong with the model.  Of course, the choice of
which probability should actually be used to draw the line between
``good'' and ``bad'' models is a purely subjective judgement.  For our
criterion, we choose that probability to be a few percent.

\subsection{Estimated Errors in the Parameters}

Assuming the model is determined to be a ``good'' one, the parameters
found at ${\chi_{\rm min}}^2$ may reveal some interesting physics.
If, for example, an unexpectedly low freezeout temperature is
discovered in a fit, it is necessary to know how confident one can be
in that particular low value.  In other words, we need to estimate the
possible error in that fitted parameter due to random measurement
errors in the experiment.  More generally, one would like to determine
the region in $M$-dimensional parameter space which has a high
probability of containing the underlying true parameter values.  For
example, we would like to be able to say ``there is a 99\%
chance that the true parameter values fall within such and such region
of parameter space.''

For normally distributed measurement errors, the relevant region is
one consisting of all combinations of parameters ${\bf\theta}$ which
would lead to a value of $\chi^2({\bf\theta})$ less than ${\chi_{\rm
min}}^2+\Delta$, where $\Delta$ is some constant number.  The
confidence level ``CL'' for a specific $\Delta$ and a specific number
of parameters $M$ is given by the integral of an $M$-degree-of-freedom
chi-square distribution \cite{numrec,stat}:
\begin{equation}
{\rm CL} = \int_0^\Delta d\chi^2 \Pi_M\!\left(\chi^2\right)
\label{6.1}
\end{equation}
For example, if we choose a region defined by $\Delta=21.666$ for a
model with $M=9$, Eq.~(\ref{6.1}) tells us that CL $=0.99$.  This
means that we have 99\% confidence that the true parameter
set lies among all possible sets which result in a
$\chi^2({\bf\theta})$ less than ${\chi_{\rm min}}^2+21.666$.  By
inverting Eq.~(\ref{6.1}), it is possible to find the appropriate
$\Delta$ for $M$ parameters which leads to any desired confidence
level.

Once $\Delta$ has been determined in this way, the desired region in
parameter space is found by making a Taylor expansion of
$\chi^2({\bf\theta})$ about its minimum at $\bar{\bf\theta}$:
\begin{equation}
\chi^2({\bf\theta}) = {\chi_{\rm min}}^2 + \left.\frac{1}{2}
\sum_{a,b=1}^M
(\theta_a-\bar{\theta}_a)(\theta_b-\bar{\theta}_b)
\frac{\partial^2\chi^2}{\partial \theta_a \partial
\theta_b}\right|_{\bar{\bf\theta}} + {\cdot}\;{\cdot}\;{\cdot}\;\;.
\label{6.2}
\end{equation}
Notice that the gradient terms disappear because we are expanding
around a minimum.  If we assume that higher-order terms are also
negligible close to the minimum, then the desired region of parameter
space is just the $M$-dimensional hyperellipsoidal volume defined by
\begin{equation}
\sum_{a,b=1}^M
(\theta_a-\bar{\theta}_a)D_{ab}(\theta_b-\bar{\theta}_b)
\le \Delta\;,
\label{6.3}
\end{equation}
where the curvature matrix $D_{ab}$ is just one-half of the second
derivative matrix of $\chi^2$.

From Eq.~(\ref{5.3}), it is apparent that the exact curvature matrix
$D_{ab}$ involves terms of the form
\[
\left[\frac{P_\alpha^i-P_\alpha({\bf p}_i,{\bf \theta})}
{{\sigma_i^{\rm stat}}^2 + (f_\alpha P_\alpha^i)^2}\right]
\frac{\partial^2P_\alpha({\bf p}_i,{\bf \theta})}
{\partial\theta_a \partial\theta_b}
\]
and
\[
\left[\frac{C_\beta^i-C_\beta({\bf q}_i,{\bf K}_i,{\bf \theta})}
{{\sigma_i}^2}\right]
\frac{\partial^2C_\beta({\bf q}_i,{\bf K}_i,{\bf \theta})}
{\partial\theta_a \partial\theta_b}\;.
\]
For a good model, $P_\alpha^i-P_\alpha$ and $C_\alpha^i-C_\alpha$ are
nothing more than the random measurement errors at the point $i$.
Since these errors can have either sign and should in general be
uncorrelated with the model, terms of the above form tend to cancel
out when summed over $i$.  By dropping these terms, we obtain the
approximate curvature matrix that is actually used in the model:
\begin{eqnarray}
D_{ab} &=& \sum_\alpha\sum_i
\frac{1}{{\sigma_i^{\rm stat}}^2 + (f_\alpha P_\alpha^i)^2}
\left.\left(\frac{\partial P_\alpha({\bf p}_i,{\bf\theta})}
{\partial\theta_a}\right)
\left(\frac{\partial P_\alpha({\bf p}_i,{\bf\theta})}
{\partial\theta_b}\right)\right|_{\bar{\bf\theta}}
\nonumber \\
&+& \sum_\beta\sum_i
\frac{1}{{\sigma_i}^2}
\left.\left(\frac{\partial C_\beta({\bf q}_i,{\bf K}_i,{\bf\theta})}
{\partial\theta_a}\right)
\left(\frac{\partial C_\beta({\bf q}_i,{\bf K}_i,{\bf\theta})}
{\partial\theta_b}\right)\right|_{\bar{\bf\theta}}\;.
\label{6.4}
\end{eqnarray}

The best and most complete way to specify the parameter confidence
region is to calculate the above curvature matrix and then employ
Eq.~(\ref{6.3}).  However, in order to get a quick idea of the size of
the region, it is useful to provide one-dimensional error estimates on
each parameter.  One way to do this is by determining the largest and
smallest values of each parameter that can be obtained on the
hyperellipsoid.  For example, suppose we had only two parameters, $T$
and $v_{\rm t}$, and we determined that the confidence region was the
interior of the ellipse shown in Fig.~3.  The one-dimensional error
estimates $\delta T$ and $\delta v_{\rm t}$ shown in the figure would
give one a rough idea of the size of the ellipse.  One must be careful
in using these one-dimensional simplifications, however, since for
example the point $(T+\delta T, v_{\rm t}+\delta v_{\rm t})$ lies
outside the confidence region.  Mathematical determination of the
one-dimensional error estimates is derived in Appendix A for the
adjusted parameters and in Appendix B for additional calculated
quantities of physical interest.

\section{Fitting to E-802 Data}

\subsection{Description of the Data}

Having developed our model, we used it to fit meson data from central
Si + Au collisions at $p_{\rm lab}/A=14.6$ GeV/$c$ measured in
experiment E-802 at the AGS \cite{e802,dbase,vince}.  As mentioned
previously, we did not attempt to fit proton or deuteron data due to
the difficulty of disentangling spectators from participants.  The
invariant one-particle multiplicity distributions for $\pi^+$,
$\pi^-$, $K^+$, and $K^-$ were obtained from an on-line database
\cite{dbase} and then organized into input files in which each line
contained the information
\[
y_{{\rm p},i}\;\;\;\;\;\;
p_{{\rm t},i}\;\;\;\;\;\;
P^i\;\;\;\;\;\;
\sigma_i^{\rm stat}\;.
\]
These data were presented in graphical form in
\cite{e802}, where it was estimated that systematic errors for all of
the particles except $K^-$'s were 10--15\%, while those for the $K^-$'s
may have been as much as 20\%.  For our fits, we used constant
systematic errors of 15\% for pions and $K^+$'s and 20\% for $K^-$'s.

In addition to these data, we obtained preliminary three-dimensional
correlation data for both $\pi^+$ and $K^+$ \cite{vince}.  For the
correlation data, the momenta of the identical particles were measured
relative to a fixed frame at $y_{\rm lab}=1.25$.  Each two-particle
event was then three-dimensionally binned according to the components
$q_z$, $q_{\rm out}$, and $q_{\rm side}$ of its momentum difference.
We use here the standard notation of $z$ denoting the beam direction,
``out'' denoting the direction parallel to the component of the
average momentum ${\bf K}$ which is perpendicular to the beam, and
``side'' denoting the remaining transverse direction\cite{bertsch2}.
The convention was used that $q_{\rm out} = p_{1,{\rm out}}-p_{2,{\rm
out}}$ is always positive \cite{chap3}.  Since this convention
identifies which particle is ``particle 1'' and which is ``particle
2,'' the remaining two momentum differences, $q_z = p_{1,z}-p_{2,z}$
and $q_{\rm side}=p_{1,{\rm side}}-p_{2,{\rm side}}$, can and did take
either sign.  In other words, there were bins with negative values of
$q_z$ and $q_{\rm side}$ as well as those with positive values.  For
each bin in ${\bf q}$, the average values of
$Y={\textstyle\frac{1}{2}}(y_1+y_2)$ and $K_{\rm t} =
{\textstyle\frac{1}{2}}(p_{1,{\rm out}}+p_{2,{\rm out}})$ for the
pairs in that bin were calculated and recorded.  Thus, in the
correlation input files, each line contained the information
\[
\langle Y\rangle_i\;\;\;\;\;\;
\langle K_{\rm t}\rangle_i\;\;\;\;\;\;
q_{z,i}\;\;\;\;\;\;
q_{{\rm out},i}\;\;\;\;\;\;
q_{{\rm side},i}\;\;\;\;\;\;
C^i\;\;\;\;\;\;
\sigma_i\;,
\]
where $\langle\rangle_i$ represents the average value over the bin $i$.

Due to the very large number of correlation data points, not all of
the data were used.  In particular, for $\pi^+$ we used only points
for which the ${\bf q}_i$ at the center of the bin satisfied the
inequality
\begin{equation}
\left(\frac{q_z}{200}\right)^2 +
\left(\frac{q_{\rm out}}{100}\right)^2 +
\left(\frac{q_{\rm side}}{100}\right)^2 < 1\;,
\label{limitpi}
\end{equation}
with the components of ${\bf q}$ measured in MeV/$c$.  Similarly, for
$K^+$ we used only data points satisfying
\begin{equation}
\left(\frac{q_z}{250}\right)^2 +
\left(\frac{q_{\rm out}}{125}\right)^2 +
\left(\frac{q_{\rm side}}{125}\right)^2 < 1\;.
\label{limitK}
\end{equation}
Our rationale for omitting some of the data points was that most of
the physically important information is contained at low momentum
differences.  For momentum differences greater than those specified
above, the model predicted a correlation function very close to one,
while the actual data fluctuated wildly about this value with huge
error bars.  Furthermore, we found that the use of different
momentum-difference cutoffs did not significantly affect our results.

Combining both the one-particle distributions and the two-particle
correlations, we used a total of 1416 data points in our fits.  Since
there are nine adjustable parameters, there are 1407 degrees of
freedom.

\subsection{Results of the Fit}

The best-fit parameter values and their estimated errors (see Appendix
A) at a 99\% confidence level for the minimum $\chi^2$ found are
listed in Table I\@.  Since the 99\%-confidence hyperellipsoid
overlaps an unphysical region in which $\alpha_{\rm t}<-1$, the lower
error estimate on $\alpha_{\rm t}$ has been reduced to reflect only
the physical region.  The curvature matrix $D$ and its inverse, the
covariance matrix $D^{-1}$, are presented in Tables II and III,
respectively.  Having found the minimum, it is possible to calculate a
number of other related quantities of physical interest.  Some of
these are constrained parameters such as $\mu_{\rm s}$, while others
are calculated quantities such as the maximum longitudinal expansion
velocity $v_\ell = {\rm tanh}(\eta_0)$ achieved by the source in its
center-of-mass frame.  One of the most interesting of these quantities
is the local baryon density at freezeout.  Appendix B describes how
each of the related quantities is calculated, while Table IV shows all
of these quantities with their estimated errors.

The ``goodness'' of the fit can be seen by looking at Fig.~4, which
shows a plot of the probability density $\Pi_\nu$ of Eq.~(\ref{7.0})
for 1407 degrees of freedom versus ${\chi_{\rm min}}^2$.  The high
probability density of the resulting ${\chi_{\rm min}}^2=1484.6$ is
good evidence that there is nothing seriously wrong with the model.
In fact, integration of the shaded region determines that there is a
7.4\% chance of obtaining a value of ${\chi_{\rm min}}^2$ at
least that large when fitting an absolutely perfect model to the data.
To get an idea of how much each data set contributed to the overall
${\chi_{\rm min}}^2$, we have broken down the individual contributions
in Table V\@.  Obviously, all of the data sets are fit reasonably
well, although the $\chi^2$ per degree of freedom for the
negative-pion one-particle distribution is slightly higher than the
rest.

The goodness of fit for the one-particle distributions can also be
seen qualitatively by looking at direct comparisons of the model to
the data.  Figures 5--8 show the theoretical and experimental meson
invariant one-particle multiplicity distributions as functions of
$m_{\rm t}-m$ for various rapidities.  As expected from the low value
of ${\chi_{\rm min}}^2$, the overall agreement looks excellent.

Although proton data were not used for the main fit, once the best
parameters have been found, it is possible to calculate the proton
distribution and compare it with the experimental data.  Figure 9
shows that the model predictions agree with the proton data moderately
well for rapidities of 1.3 or greater.  For lower rapidities, however,
the data show far more low-$p_{\rm t}$ protons than are predicted by
the model.  In our picture, we consider these protons to be target
spectators which may have interacted somewhat, but not enough to be
considered part of the hydrodynamical system.  It should also be noted
that since our model does not distinguish between a deuteron and a
separate proton and neutron, some of the excess when the model
overpredicts the proton data may be due to leaving out deuteron
coalescence.

As mentioned before, the main problem with including proton data in
the fits is figuring out how to eliminate contamination by spectators.
Nevertheless, in order to get some idea of how the inclusion of
protons might affect our results, we made four different fits in which
we included all proton data in the following rapidity ranges: $1.5\le
y_{\rm p}\le 1.9$, $1.3\le y_{\rm p}\le 1.9$, $1.5\le y_{\rm p}\le
2.1$, and $1.3\le y_{\rm p}\le 2.1$.  These fits all give extremely
similar results, so we will discuss only the last case.  The 190
proton data points for this case bring the number of degrees of
freedom up to 1597.  The minimum $\chi^2$ of 1730.4
($\chi^2/\nu=1.084$) is obtained with the following parameters:
$T=95.8\pm 3.9$ MeV, $\mu_{\rm b}/T=5.30\pm 0.28$,
$\lambda_\pi=0.66\pm 0.11$, $R=7.8\pm 1.5$ fm, $v_{\rm t}=0.640\pm
0.034\;c$, $\alpha_{\rm t}=-$0.87 $^{+0.35}_{-0.13}$, $y_{\rm
s}=1.320\pm 0.072$, $\eta_0=1.48\pm 0.14$, and $\tau_{\rm f}=8.1\pm
2.1$ fm/$c$.  The central values themselves for all of the parameters
except $\mu_{\rm b}/T$ lie within the individual 99\%-confidence
intervals of the original fit (see Table I).  Even for $\mu_{\rm
b}/T$, the 99\%-confidence interval for the fit with protons overlaps
that for the original fit.  In other words, inclusion of the proton
data changes the best-fit parameter values only within their stated
uncertainties.  It does, however, lead to a significant reduction in
the calculated number of projectile participants.  For the fit with
protons included, the number of projectile participants is reduced to
$17.7\pm 1.4$, which is to be compared to 26.1 $^{+8.8}_{-6.7}$ found
in Table IV and to 28 nucleons in a $^{28}$Si nucleus.  If one were to
take the proton fit seriously, one might wonder how a central Si + Au
collision could possibly give rise to 10 projectile spectators.  It is
our view that the unresolved issues of coalescence and spectator
separation in our model make it better to simply neglect the protons
altogether, just as we did in our original fit.  For the remainder of
the paper, we will always refer only to that original fit.

Since the preliminary correlation data that was used has not yet been
published, we show here in Figs.~10 and 11 $(q_z,q_{\rm out})$
projections of the correlation functions calculated by the model.
Notice that whereas the kaon correlation function intercepts the ${\bf
q}=0$ axis at 2, the pion correlation function intercepts the axis at
1.65, corresponding to a value of $\lambda_\pi=0.65$.  Notice also
that since nonvanishing values of both $K_{\rm t}$ and $Y-y_{\rm s}$
are used for the plots, the effects of the ``out-long'' cross term
\cite{chap2,chap4,chap5} 
can be seen (especially in the kaons) as a slight twisting
in the major axes of the correlation function.

Data from central Si + Au collisions at the AGS have also been
compared to a thermal model in \cite{bm}.  There it was argued that a
freezeout temperature of 120--140 MeV was consistent with these data.
We, on the other hand, have found a much lower temperature of $92.9\pm
4.4$ MeV (see Table I).  To explain this significant discrepancy, we
would like to point out a few differences between our approach and
that of \cite{bm}.  First of all, they assumed that transverse and
longitudinal flow are completely separable.  This led them to compare
a thermal model that had been integrated over all rapidities
\cite{schn} with data from a single mid-rapidity bin.  In contrast, 
we make no such assumption.  Secondly, the model used in \cite{bm}
never specifies the size or shape of the freezeout hypersurface.  In
addition to preventing comparisons to two-particle correlations, this
ambiguity forces them to multiply each of their one-particle
distributions by an arbitrary normalization factor before comparing to
the normalized data of \cite{e802}.  Not only does our unambiguous
parametrization of the freezeout hypersurface allow us to compare to
two-particle correlations, it also allows us to see the significant
effects that different temperatures have on the absolute
normalizations of one-particle distributions.  Another significant
difference between the two approaches is that in
\cite{bm} only two points in temperature were studied, whereas in our
approach the whole nine-dimensional parameter space is explored,
resulting in the absolute minimum of $\chi^2$.

In order to see how much worse higher temperature fits would be, we
made a number of runs at various fixed temperatures, allowing all of
the other parameters to vary.  The results are given in Table VI and
in Fig.~12, a plot of the minimum $\chi^2$ at a fixed temperature vs.\
that temperature.  As Table VI shows, the value of $\chi^2$ for
$T=120$ MeV is 2025.4.  Again by integrating $\Pi_\nu$ of
Eq.~(\ref{7.0}), we can determine that the probability of a perfect
model resulting in a $\chi^2$ at least as large as 2025.4 is the
incredibly small value $5.1\times 10^{-25}$.  Above $T=129$ MeV, the
minimum $\chi^2$ solution switches to a different branch.  This
high-temperature branch is actually unphysical, as can be seen by
examining Table VI and noting that it is impossible for the system to
expand to the large transverse radius $R$ in the infinitesimally
small time $t_1=\tau_{\rm f}\sqrt{1+\alpha_{\rm t}}$.

Another result of the model is the size and shape of the freezeout
hypersurface.  Since $\alpha_{\rm t}$ for the best fit is negative,
freezeout begins at time $t=t_1=3.1$ fm/$c$ at $z=0$ and $\rho=R=8.0$
fm, and takes 5.1 fm/$c$ to reach the center of the source at
$z=\rho=0$ at time $t=t_2=\tau_{\rm f}=8.2$ fm/$c$.  Freezeout along
the symmetry axis then occurs at a constant proper time, finally
ending at source-frame time $t_3=\tau_{\rm f}\cosh\eta_0=18.8$ fm/$c$.
As mentioned previously, Figs.~1 and 2 pictorially show the freezeout
process for these parameter values.

\section{Future Issues}
The actual ``freezeout'' process taking place in these collisions is
undoubtedly far more complicated than in our model.  Azimuthal
symmetry may be broken, the local temperature and chemical potentials
may have some spacetime dependence, the expansion flow velocity may be
neither boost-invariant nor linear in $\rho$, the hypersurface may
have a different shape and/or some four-dimensional fuzziness, kaons
may freeze out before pions, chemical equilibrium may not be fully
achieved, etc.  One may even question whether equilibrium
hydrodynamical concepts are valid at all.  Although this paper does
not definitively settle these questions, the remarkable agreement
between theory and experiment suggests that our realistic
nine-parameter expanding source model nevertheless provides a very
good description of the most important physics taking place at
freezeout.

One parameter which definitely needs to be better understood is the
incoherence parameter $\lambda_\pi$.  Does the fact that it is less
than one mean that a significant number of pions are being produced
coherently, or could the reduced intercept instead be largely an
artifact arising from the way the correlation function was determined
experimentally
\cite{jacak}?  

We hope that our model will be used in the future to systematically
analyze the dependence of the freezeout quantities upon bombarding
energy and the sizes of the colliding nuclei.  A sharp discontinuity
in one or more of these quantities could be a signal of
quark-gluon-plasma formation.

\acknowledgments We are grateful to Arnold J. Sierk for his
participation in the early stages of this work, to Bernd R. Schlei for
illuminating discussions about hydrodynamics and freezeout
hypersurfaces, and to T. Vincent A.  Cianciolo for permitting us to
use his preliminary data on two-particle correlations in our
adjustments.  This work was supported by the U. S. Department of
Energy.

\appendix

\section{One-Dimensional Error Projections}

The simplest way to determine the largest and smallest values attained
by parameter $\theta_a$ on the hyperellipsoid defined by
Eq.~(\ref{6.3}) is to use a Lagrange multiplier $\xi_a$.  We begin by
finding the maximum (or minimum) of the quantity
\[
\theta_a - \xi_a\sum_{b,c=1}^M
(\theta_b-\bar{\theta}_b)D_{bc}(\theta_c-\bar{\theta}_c)\;.
\]
By differentiating with respect to $\theta_d$, we find the coordinates
$\theta_e$ of the extrema as a function of $\xi_a$:
\begin{equation}
\theta_e-\bar{\theta}_e = \frac{1}{2\xi_a}\left(D^{-1}\right)_{ea}\;,
\label{a1}
\end{equation}
where the subscripts $a$ are not summed over.  To impose the
constraint that the solution lies on the hyperellipsoid, we must pick
a $\xi_a$ such that the equality of Eq.~(\ref{6.3}) is satisfied.
Plugging Eq.~(\ref{a1}) into Eq.~(\ref{6.3}) and solving for $\xi_a$,
we find
\begin{equation}
2\xi_a = \pm\sqrt{\frac{\left(D^{-1}\right)_{aa}}{\Delta}}\;,
\label{a2}
\end{equation}
where again the indices $a$ are not summed over.  Inserting
Eq.~(\ref{a2}) into Eq.~(\ref{a1}), we find the values of all of the
parameters $\theta_e$ corresponding to each extremum of $\theta_a$ on
the hyperellipsoid.  In particular, the one-dimensional error estimate
on $\theta_a$ is just proportional to the square root of the $aa$th
element of the covariance matrix (the inverse of the curvature
matrix):
\begin{equation}
\theta_a-\bar{\theta}_a = \pm\sqrt{\Delta\left(D^{-1}\right)_{aa}}\;.
\label{a3}
\end{equation}
By inverting the curvature matrix to get the covariance matrix, one
can then just read off the one-dimensional error estimate on each
parameter by taking the square root of the product of the appropriate
diagonal element times $\Delta$.

\section{Additional Calculated Quantities}

Here we list some additional physical quantities of interest which can
be calculated from the nine parameters.  From Eqs.~(\ref{2.4}) and
(\ref{2.11}), it can be seen that the maximum longitudinal velocity
achieved by the source is given by $v_\ell=\tanh\eta_0$.  Also, in
Sec.~II.B the times that freezeout begins, reaches the center of the
source, and ends were shown to be $t_1=\tau_{\rm f}\sqrt{1+\alpha_{\rm
t}}$, $t_2=\tau_{\rm f}$, and $t_3=\tau_{\rm f}\cosh\eta_0$,
respectively.  The maximum longitudinal extension of the source is
given by $z_3=\tau_{\rm f}\sinh\eta_0$, while the duration of
freezeout at $z=0$ is given by $\Delta t=t_2-t_1$.  The total baryon
number $B_{\rm tot}$ of the source is given by the denominator of the
second equation in (\ref{3.2}).  Section II.C also explains how the
chemical potentials $\mu_{\rm s}$ and $\mu_{\rm i}$ are found.  The
numbers of projectile and target participants can be deduced from
Eq.~(\ref{3.5}) and are given by
\begin{eqnarray}
B_{\rm proj} &=& \frac{B_{\rm tot}\sinh(y_{\rm s}-y_{\rm tar})}
{\sinh(y_{\rm s}-y_{\rm tar}) + \sinh(y_{\rm proj}-y_{\rm s})}
\nonumber \\
B_{\rm tar} &=& \frac{B_{\rm tot}\sinh(y_{\rm proj}-y_{\rm s})}
{\sinh(y_{\rm proj}-y_{\rm s}) + \sinh(y_{\rm s}-y_{\rm tar})}\;.
\label{b1}
\end{eqnarray}

The local density of particles of type $\alpha$ is given by the
integral 
\begin{eqnarray}
n_\alpha(\rho) &=& \frac{1}{\gamma_\rho\cosh\eta}
\int\frac{d^3p}{E}\int dt S(x,p)\;.
\nonumber \\
&=& \frac{J_\alpha(J_\alpha+1)}{2\pi^2\gamma_\rho}
\sum_{k=1}^\infty(-1)^k\exp(k\mu_\alpha/T)
\int m_{\rm t}dm_{\rm t}
\left[m_{\rm t}K_1\left(\frac{k\gamma_\rho m_{\rm t}}{T}\right)
I_0\left(\frac{K\gamma_\rho p_{\rm t} v_{\rm t}\rho}{RT}\right)\right.
\nonumber \\
&&-\left.\frac{\alpha_{\rm t}p_{\rm t}\rho\tau_{\rm f}}
{R^2\sqrt{1+\alpha_{\rm
t}(\rho/R)^2}} 
K_0\left(\frac{k\gamma_\rho m_{\rm t}}{T}\right)
I_1\left(\frac{K\gamma_\rho p_{\rm t} v_{\rm t}\rho}{RT}\right)
\right]\;,
\label{b2}
\end{eqnarray}
where $K_i$ and $I_i$ are modified Bessel functions of order $i$.  Due
to the boost invariance assumed in everything but the spatial limits
of the model, $n_\alpha$ is a function of only $\rho$ and not of
$\eta$.  The local baryon density is just given by
\begin{equation}
n_{\rm b}(\rho) = \sum_\alpha B_\alpha n_\alpha(\rho)\;.
\label{b3}
\end{equation}

To calculate error estimates on these quantities, we could in
principle use a more general form of the Lagrange-multiplier method
introduced in Appendix A\@.  We have found, however, that a quicker
and more reliable method is to first find $M-1$ new variables which
can parametrize just the surface of the hyperellipsoid, express the
quantities as functions of these new variables, and then find the
extrema of these functions.  We begin this process by numerically
finding the unitary matrix $U$ which transforms $D$ into the diagonal
matrix $\tilde{D}$, namely
\begin{equation}
\tilde{D} = U^{-1}\,D\,U\;.
\label{c1}
\end{equation}
Next we use the (diagonal) elements of $\tilde{D}$ to define the
vector 
\begin{equation}
\psi_a = \sqrt{\frac{\tilde{D}_{aa}}{\Delta}}
\sum_{b=1}^M \left(U^{-1}\right)_{ab}(\theta_b-\bar{\theta}_b)\;,
\label{c2}
\end{equation}
where there is no summation over the index $a$.  Using these
variables, we can see that the equality of Eq.~(\ref{6.3}) reduces to
the equation of the surface of a sphere in $M$ dimensions, namely
\begin{equation}
\sum_{a=1}^M {\psi_a}^2 = 1\;.
\label{c3}
\end{equation}
Since the surface is ($M-1$)-dimensional, $M-1$ angles are sufficient to
identify any point on it.  We now redefine the $\psi_a$ in terms of
the angles $\phi_a$ through
\begin{eqnarray}
\psi_1 &=& \cos(\phi_1)
\nonumber \\
\psi_2 &=& \sin(\phi_1)\cos(\phi_2)
\nonumber \\
&\vdots&
\nonumber \\
\psi_{M-1} &=& \sin(\phi_1)\sin(\phi_2)
\cdots\sin(\phi_{M-2})\cos(\phi_{M-1})
\nonumber \\
\psi_M &=& \sin(\phi_1)\sin(\phi_2)
\cdots\sin(\phi_{M-2})\sin(\phi_{M-1})\;.
\label{c4}
\end{eqnarray}
Since the inverse of Eq.~(\ref{c2}) tells us
\begin{equation}
\theta_a = \bar{\theta}_a + \sqrt{\frac{\Delta}{\tilde{D}_{aa}}}
\sum_{b=1}^MU_{ab}\psi_b\;,
\label{c5}
\end{equation}
we now have the $M$ parameters $\theta_a$ expressed in terms of the
$M-1$ parameters $\phi_a$.  As mentioned previously, the extrema on
the hyperellipsoid for any function $f({\bf\theta}({\bf\phi}))$ can be
found simply by allowing the $\phi_a$ to vary freely.

%%%%%%%%%%%%%%%%%%%%%%%%%%%%%%%%%%%%%%%%%%%%%%%%%%%%%%%%%%%%%%%%%
%\begin{thebibliography}{99}

%\end{thebibliography}

%%%%%%%%%%%%%%%%%%%%%%%%%%%%%%%%%%%%%%%%%
%  FIGURES
%%%%%%%%%%%%%%%%%%%%%%%%%%%%%%%%%%%%%%%%%

\begin{figure}
\caption{Freezeout hypersurface, 
which specifies the positions in spacetime where the expanding
hydrodynamical fluid is converted into a collection of noninteracting,
free-streaming hadrons.  The parameters used are $R=8.0$ fm,
$\tau_{\rm f}=8.2$ fm/$c$, $\eta_0=1.47$, and $\alpha_{\rm t}=-0.86$.}
\label{1}
\end{figure}
\begin{figure}
\caption{Seven snapshots at equal spacings in the source-frame time 
$t$ of the part of the source of Fig.~\ref{1} that is freezing out
(inner surfaces) or has not yet frozen out (end caps).  To draw the
end caps at each time $t$, we used $v_{\rm t}=0.683$ $c$ and assumed
that none of the fluid was accelerated prior to reaching the freezeout
hypersurface.}
\label{2}
\end{figure}
\begin{figure}
\caption{An example 99\%-confidence ellipse for a model with
two parameters showing how one-dimensional errors on those parameters
are estimated.}
\label{3}
\end{figure}
\begin{figure}
\caption{The probability density $\Pi_\nu$ of Eq.~(\ref{7.0}) for
obtaining a particular minimum of $\chi^2$ from a fit with 1407
degrees of freedom $\nu$ to a perfect model.  There is a 7.4\% chance
that a perfect model would have given rise to a ${\chi_{\rm min}}^2$
at least as large as the one actually found, 1484.6.}
\label{4}
\end{figure}
\begin{figure}
\caption{Comparison between model predictions and experimental data
[6,7] for the invariant $\pi^+$ one-particle multiplicity distribution
$Ed^3N/dp^3=1/(2\pi m_{\rm t})\,d^2N/dy_{\rm p}dm_{\rm t}$ as a
function of $y_{\rm p}$ and $m_{\rm t}-m$.  For visual separation, the
results at a particular particle rapidity $y_{\rm p}$ relative to the
laboratory frame are scaled by the factor $10^{5(0.7-y_{\rm p})}$.
Although not always distinguishable on the scale of the graphs,
statistical errors are given by the inner error bars, and total errors
are given by the outer error bars in \protect\vspace*{-12.7pt} Figs.\
5--9.}
\label{5}
\end{figure}
\begin{figure}
\caption{Comparison between model and experimental
$\pi^-$ distributions, scaled as in \protect\vspace*{-12.7pt}
Fig.~\ref{5}.}
\label{6}
\end{figure}
\begin{figure}
\caption{Comparison between model and experimental
$K^+$ distributions, scaled as in \protect\vspace*{-12.7pt}
Fig.~\ref{5}.}
\label{7}
\end{figure}
\begin{figure}
\caption{Comparison between model and experimental
$K^-$ distributions, scaled as in Fig.~\ref{5}.}
\label{8}
\end{figure}
\begin{figure}
\caption{Comparison between model and experimental
proton distributions, scaled as in Fig.~\ref{5}.  
Note that these proton data were not included in the fit.}
\label{9}
\end{figure}
\begin{figure}
\caption{Dependence of the predicted $\pi^+$ two-particle correlation
function $C$ upon the longitudinal and ``out'' momentum differences, for
fixed values of the other three quantities upon which $C$ depends.}
\label{10}
\end{figure}
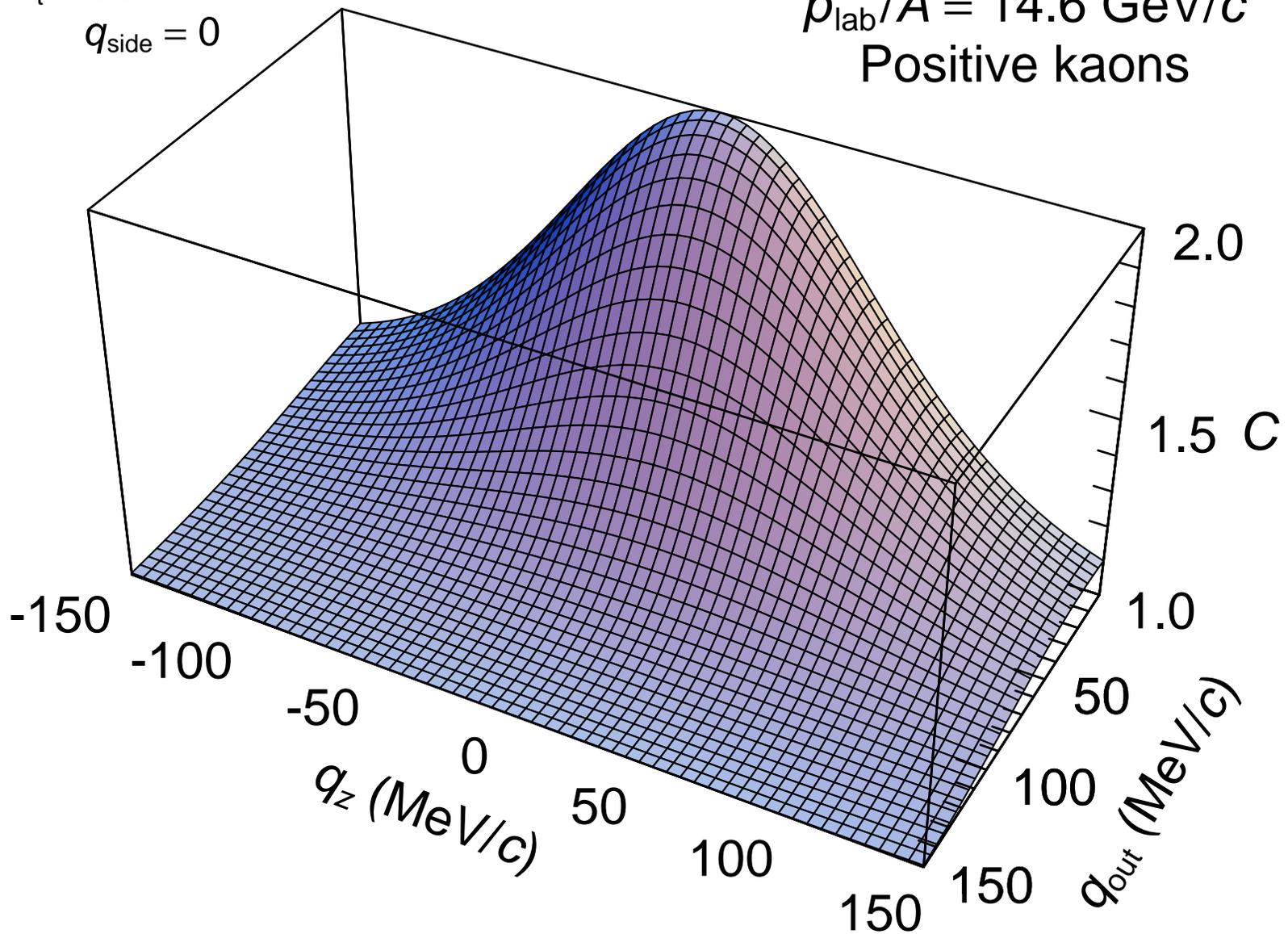
\begin{figure}
\caption{Dependence of the predicted $K^+$ two-particle correlation
function $C$ upon the longitudinal and ``out'' momentum differences,
for fixed values of the other three quantities upon which $C$
depends.}
\label{11}
\end{figure}
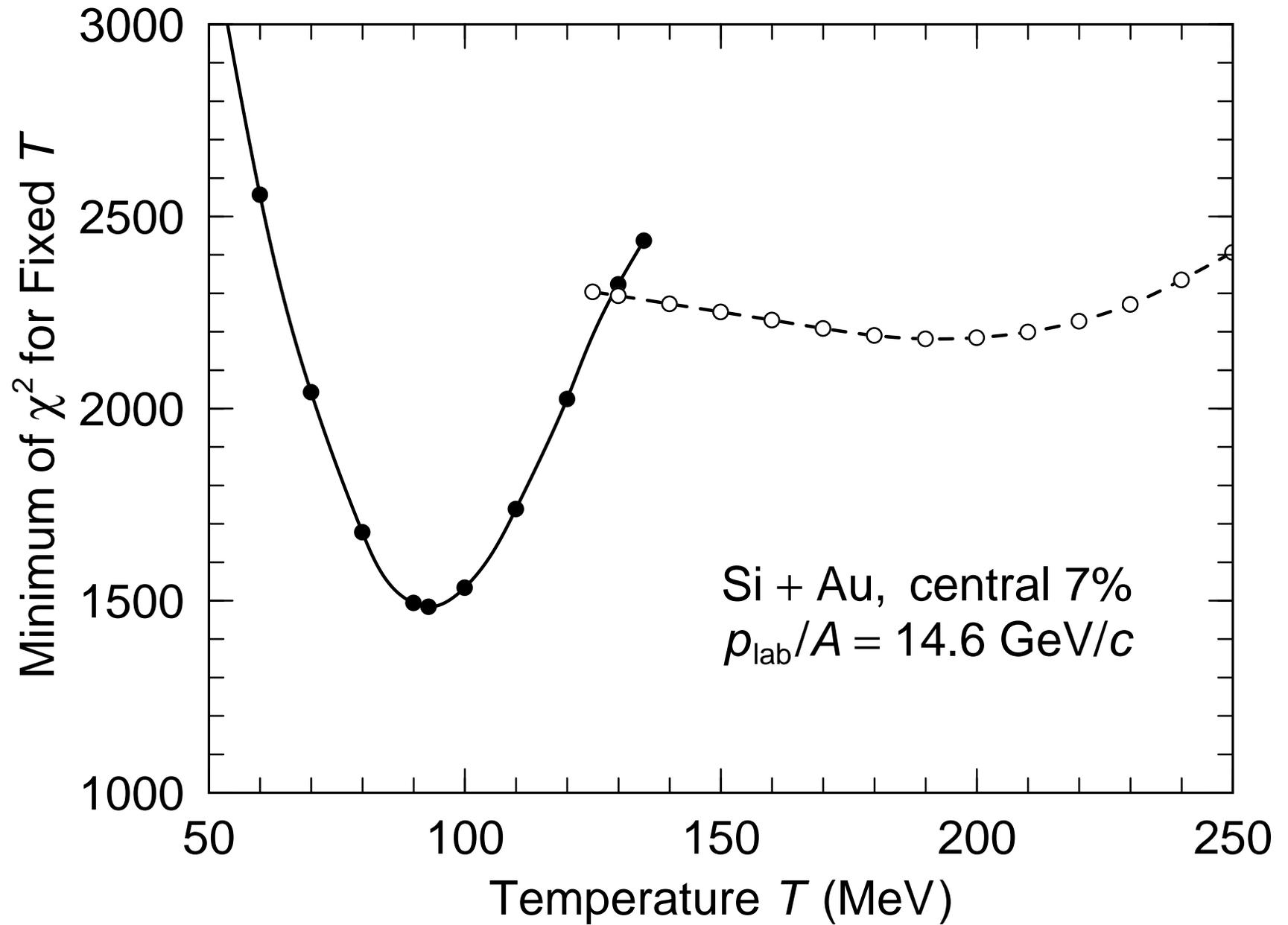
\begin{figure}
\caption{Minima of $\chi^2$ for fixed values of $T$, when all other
parameters are allowed to vary.  Solid circles connected by a solid
line identify one branch which contains some physically relevant
solutions near the minimum, while open circles connected by a dashed
line represent a different branch containing only unphysical solutions
(see Table VI).}
\label{12}
\end{figure}

%%%%%%%%%%%%%%%%%%%%%%%%%%%%%%%%%%%%%%%%%
%  TABLES
%%%%%%%%%%%%%%%%%%%%%%%%%%%%%%%%%%%%%%%%%

\begin{table}
\caption{Nine adjusted source freezeout parameters.}
{\begin{tabular}{lc}  

& Value and uncertainty \\
 
Parameter & at 99\% confidence \\ \hline

Nuclear temperature $T$ & 92.9 $\pm$ 4.4 MeV \\

Baryon chemical potential $\mu_{\rm b}/T$ & 5.97 $\pm$ 0.56 \\

Pion incoherence fraction $\lambda_{\pi}$ & 0.65 $\pm$ 0.11 \\

Transverse freezeout radius $R$ & 8.0 $\pm$ 1.6 fm \\

Transverse freezeout velocity $v_{\rm t}$ & 0.683 $\pm$ 0.048 $c$ \\

Transverse freezeout coefficient $\alpha_{\rm
t}$\tablenote{Physically, $\alpha_{\rm t}$ cannot be less than $-1$,
since that value corresponds to freezeout beginning immediately at
$t=0$.} & $-$0.86 $^{+0.37}_{-0.14}$ \\

Source rapidity $y_{\rm s}$ & 1.355 $\pm$ 0.066 \\

Longitudinal spacetime rapidity $\eta_0$ & 1.47 $\pm$ 0.13 \\

Longitudinal freezeout proper time $\tau_{\rm f}$ & 8.2 $\pm$ 2.2
fm/$c$ \\
\end{tabular}} 
\end{table}

\begin{table}
\caption{Curvature matrix $D$.  Derivatives with respect to
the parameters are ordered in rows and columns in the same way as for
the parameters in Table I\@.  The units of each element are given by
the inverse of the units of the associated row and column parameters
(e.g., the units of $D_{14}$ are 1/(MeV${\cdot}$fm)).}
\begin{tabular}{ccrrrrrrrrr}
& & 65.72 & 117.4 & \hspace{.13cm} $-$11.16 &
175.6 & 1859 & \hspace{.13cm} $-$54.69 &
606.7 & 524.4 & 140.5 \\
& & 117.4 & 393.1 & $-$9.373 & 295.8 & 1717 & $-$36.53 & 867.3
& 740.1 & 222.9 \\
& & $-$11.16 & \hspace{.13cm} $-$9.373 & 2017 &
\hspace{.13cm} $-$127.6 & 368.7 & 100.0 &
\hspace{.13cm} $-$110.7 & \hspace{.13cm} $-$124.2 & \hspace{.13cm}
$-$82.88 \\ & & 175.6 & 295.8 & $-$127.6 & 509.4 & 4641 & $-$169.9 &
1761 & 1501 & 395.3
\\
$D$ & $=$ & 1859 & 1717 & 368.7 & 4641 & 87117 & $-$2530 & 13964 &
14133 & 4107
\\
& & $-$54.69 & $-$36.53 & 100.0 & $-$169.9 & \hspace{.13cm} $-$2530 &
313.5 & $-$678.5 & $-$580.9 & $-$123.0
\\
& & 606.7 & 867.3 & $-$110.7 & 1761 & 13964 & $-$678.5 & 19418 & 10296
& 1315
\\
& & 524.4 & 740.1 & $-$124.2 & 1501 & 14133 & $-$580.9 & 10296 & 7643
& 1158
\\
& & 140.5 & 222.9 & $-$82.88 & 395.3 & 4107 & $-$123.0 & 1315 & 1158 &
319.0 \\
\end{tabular}
\end{table}

\begin{table}
\caption{Covariance matrix $D^{-1}$.  The ordering of the
rows and columns is the same as in Table II\@.  The units of each
element are $10^4$ times the inverse of the units in Table II (e.g.,
the units of $(D^{-1})_{14}/10^4$ are MeV${\cdot}$fm).}
\begin{tabular}{ccrrrrrrrrr}
& & 8860 & \hspace{.01cm} $-$891.5 & $-$53.42 & $-$1180 & $-$61.21 &
$-$121.3 & $-$11.63 & $-$29.23 & \hspace{.01cm} $-$935.8
\\
& & $-$891.5 & 143.9 & 2.678 & 76.55 & 7.407 & 5.523 & 1.195 & 4.962 &
81.77 
\\
& & $-$53.42 & 2.678 & 5.827 & \hspace{.01cm} 6.837 &
\hspace{.01cm} $-$0.04256 & \hspace{.01cm} $-$1.918 &
\hspace{.01cm} 0.04292 &
\hspace{.01cm} $-$0.3064 & 15.44 
\\
& & $-$1180 & 76.55 & 6.837 & 1244 & 24.25 & 146.0 & 1.241 &
$-$8.943 & $-$1302 
\\ 
$D^{-1}$ & $=$ & $-$61.21 & 7.407 & $-$0.04256 & 24.25 & 1.080 & 4.091
& 0.1796 & 0.04903 & $-$21.53
\\ 
& & $-$121.3 & 5.523 & $-$1.918 & 146.0 & 4.091 & 61.51 & 0.1794 &
0.9064 & $-$164.8
\\ 
& & $-$11.63 & 1.195 & \hspace{.01cm} 0.04292 & 1.241 & 0.1796 &
0.1794 & 1.998 & $-$3.112 & 3.576
\\ 
& & $-$29.23 & 4.962 & $-$0.3064 & $-$8.943 & 0.04903 & 0.9064 &
$-$3.112 & 8.359 & 2.626
\\ 
& & $-$935.8 & 81.77 & 15.44 & $-$1302 & $-$21.53 & $-$164.8 & 3.576 &
2.626 & 2193 \\
\end{tabular}
\end{table}

\begin{table}
\caption{Additional calculated physical quantities.}
\begin{tabular}{lc}  
& Value and uncertainty \\
 
Quantity & at 99\% confidence \\ \hline

Source velocity $v_{\rm s}$ & 0.875 $^{+0.015}_{-0.016}$ $c$  \\

Longitudinal velocity $v_{\ell}$ & 0.900 $^{+0.023}_{-0.029}$ $c$ \\

Longitudinal freezeout radius $z_3$ & 16.9 $^{+5.6}_{-4.9}$ fm \\

Beginning freezeout time $t_{1}$ & 3.1 $^{+2.5}_{-3.1}$ fm/$c$ \\

Freezeout time $t_{2}$ at source center & 8.2 $\pm$ 2.2 fm/$c$ \\

Final freezeout time $t_{3}$ & 18.8 $^{+5.8}_{-5.3}$ fm/$c$ \\

Freezeout width $\Delta\tau$ in proper time\tablenote{Calculated under
the additional assumption that the exterior matter at $z = 0$ that
freezes out first has been moving with constant transverse velocity
$v_{\rm t}$ from time $t = 0$ until time $t_1$.} & 5.9
$^{+4.4}_{-2.6}$ fm/$c$ \\

Baryon chemical potential $\mu_{\rm b}$ & 554 $^{+34}_{-36}$ MeV \\

Strangeness chemical potential $\mu_{\rm s}$ & 75 $^{+13}_{-12}$ MeV \\

Isospin chemical potential $\mu_{\rm i}$ & $-5.3$ $^{+1.0}_{-1.1}$ MeV \\

Number $B_{\rm proj}$ of baryons originating from projectile & 26.1
$^{+8.8}_{-6.7}$ \\

Number $B_{\rm tar}$ of baryons originating from target & 57
$^{+20}_{-15}$ \\

Total number $B_{\rm tot}$ of baryons in source & 83 $^{+28}_{-21}$ \\

Baryon density $n_1$ at beginning of freezeout\tablenote{The upper
limit of $\infty$ for this quantity arises because the beginning
freezeout time $t_1$ could be zero, at which time the shape is an
infinitesimally thin disk.}
& 0.057 $^{+\infty}_{-0.032}$ fm$^{-3}$
\\

Baryon density $n_{\rm s}$ along symmetry axis & 0.0222
$^{+0.0096}_{-0.0069}$ fm$^{-3}$ \\
\end{tabular}
\end{table}

\begin{table}
\caption{Individual contributions to $\chi^2$.}
\begin{tabular}{lrrrr} 
Type of data & $N_{\rm data}$ & $\nu$\tablenote{In
determining the number of degrees of freedom for the individual
contributions, we have allocated the nine adjustable parameters among
each type of data in proportion to the number of data points.} 
& $\chi^2$ & $\chi^2/\nu$ \\ \hline
$\pi^+$ one-particle & 231 & $\;\;\;$ 229.5 
& $\;\;\;$ 238.0 & $\;\;\;$ 1.037 \\
$\pi^-$ one-particle & 239 & 237.5 & 266.2 & 1.121 \\
$K^+$ one-particle & 137 & 136.1 & 140.6 & 1.033 \\
$K^-$ one-particle & 49 & 48.7 & 51.2 & 1.051 \\
$\pi^+$ correlation & 464 & 461.1 & 498.0 & 1.080 \\
$K^+$ correlation & 296 & 294.1 & 290.7 & 0.988 \\ \hline
Total & 1416 & 1407.0 & 1484.6 & 1.055 \\
\end{tabular}
\end{table}

\begin{table}
\caption{Eight adjusted parameters of best fits at fixed values of
temperature $T$ plus calculated $B_{\rm proj}$.  Physically relevant
solutions correspond only to a limited region surrounding the absolute
minimum at $T=92.9$ MeV\@.  For values of $T$ below 82 MeV\@, the
calculated lower limit on $B_{\rm proj}$ exceeds the number of
nucleons in the projectile.  Solutions below the horizontal line
correspond to a new branch which is unphysical for the reason
mentioned in the text.}
\begin{tabular}{ccccccccccc}
$\chi^2$ & $T$ & $\mu_{\rm b}/T$ & $\lambda_\pi$ & $R$ & $v_{\rm t}$ &
$\alpha_{\rm t}$ & $y_{\rm s}$ & $\eta_0$ & $\tau_{\rm f}$ & $B_{\rm
proj}$
\\
& (MeV) & & & (fm) & ($c$) & & & & (fm/$c$) & 
\\ \hline
3280.7 & 50 & 16.4 & 1.02 & 21.1 & 0.971 & 0.232 & 1.467 & 1.75 & 10.2
& 566
\\[-4.8pt]
2557.2 & 60 & 12.7 & 0.868 & 14.2 & 0.920 & $-0.523$ & 1.428 & 1.70 &
10.8 & 204
\\[-4.8pt]
2043.2 & 70 & 9.98 & 0.777 & 11.1 & 0.854 & $-0.671$ & 1.394 & 1.64 &
10.5 & 88.9
\\[-4.8pt]
1678.9 & 80 & 7.76 & 0.723 & 9.64 & 0.778 & $-0.738$ & 1.375 & 1.55 &
9.65 & 41.9
\\[-4.8pt]
1494.6 & 90 & 6.28 & 0.667 & 8.39 & 0.705 & $-0.818$ & 1.361 & 1.49 &
8.47 & 27.8
\\[-4.8pt]
1534.4 & 100 & 5.35 & 0.613 & 7.07 & 0.631 & $-0.999$ & 1.341 & 1.46 &
7.55 & 23.6
\\[-4.8pt]
1739.1 & 110 & 4.68 & 0.544 & 6.13 & 0.568 & $-0.999$ & 1.315 & 1.47 &
6.44 & 21.6
\\[-4.8pt]
2025.4 & 120 & 4.17 & 0.486 & 5.24 & 0.502 & $-0.999$ & 1.284 & 1.49 &
5.69 & 20.1
\\[-4.8pt]
2323.9 & 130 & 3.81 & 0.473 & 3.42 & 0.383 & $-0.999$ & 1.266 & 1.54 &
7.40 & 19.2
\\[-4.8pt]
2437.8 & 135 & 3.70 & 0.545 & 2.37 & 0.281 & $-0.999$ & 1.268 & 1.57 &
10.3 & 19.3
\\[-4.8pt] \hline
2304.5 & 125 & 7.17 & 0.530 & 13.5 & 0.872 & $-0.999$ & 1.380 & 1.47 &
0.133 & 72.5
\\[-4.8pt]
2294.2 & 130 & 7.05 & 0.530 & 13.3 & 0.866 & $-0.999$ & 1.379 & 1.47 &
0.102 & 68.2
\\[-4.8pt]
2273.1 & 140 & 6.85 & 0.529 & 13.0 & 0.854 & $-0.999$ & 1.375 & 1.45 &
0.0630 & 60.3
\\[-4.8pt]
2252.2 & 150 & 6.69 & 0.528 & 12.7 & 0.839 & $-0.999$ & 1.371 & 1.43 &
0.0407 & 53.2
\\[-4.8pt]
2230.9 & 160 & 6.60 & 0.528 & 12.4 & 0.823 & $-0.999$ & 1.268 & 1.41 &
0.0269 & 45.8
\\[-4.8pt]
2209.3 & 170 & 6.54 & 0.530 & 12.1 & 0.804 & $-0.999$ & 1.365 & 1.39 &
0.0183 & 37.5
\\[-4.8pt]
2190.9 & 180 & 6.41 & 0.532 & 11.9 & 0.783 & $-0.999$ & 1.364 & 1.36 &
0.0139 & 31.1
\\[-4.8pt]
2182.1 & 190 & 6.20 & 0.533 & 11.6 & 0.760 & $-0.999$ & 1.361 & 1.35 &
0.0116 & 28.4
\\[-4.8pt]
2185.2 & 200 & 5.96 & 0.536 & 11.4 & 0.738 & $-0.999$ & 1.359 & 1.33 &
0.0103 & 27.7
\\[-4.8pt]
2199.9 & 210 & 5.76 & 0.537 & 11.1 & 0.713 & $-0.999$ & 1.353 & 1.32 &
0.00905 & 26.4
\\[-4.8pt]
2228.1 & 220 & 5.57 & 0.537 & 10.9 & 0.690 & $-0.999$ & 1.347 & 1.32 &
0.00805 & 27.0
\\[-4.8pt]
2271.9 & 230 & 5.28 & 0.542 & 10.7 & 0.666 & $-0.999$ & 1.333 & 1.33 &
0.00765 & 26.7
\\[-4.8pt]
2335.3 & 240 & 4.99 & 0.544 & 10.5 & 0.637 & $-0.999$ & 1.325 & 1.34 &
0.00756 & 28.1
\\[-4.8pt]
2407.0 & 250 & 4.73 & 0.545 & 10.2 & 0.605 & $-0.999$ & 1.319 & 1.34 &
0.00748 & 28.7
\\
\end{tabular}
\end{table}

\end{document}